%% file: main.tex
\documentclass[sigconf]{acmart}

\usepackage[utf8]{inputenc}
\usepackage{graphicx}
\usepackage{array}
\usepackage{tabu}
\usepackage{float}
\usepackage{xcolor}
\usepackage{mathtools}
\usepackage{multirow}
\usepackage{multicol}
\usepackage{graphics}
\usepackage{xcolor}
\usepackage{subfigure}
\usepackage{caption}
\usepackage[utf8]{inputenc}
\usepackage[T1]{fontenc}
\usepackage{fancyhdr}
\usepackage{listings}
\usepackage{xcolor}
\renewcommand\footnotetextcopyrightpermission[1]{} 
\definecolor{codegreen}{rgb}{0,0.6,0}
\definecolor{codegray}{rgb}{0.5,0.5,0.5}
\definecolor{codepurple}{rgb}{0.58,0,0.82}
\definecolor{backcolour}{rgb}{0.95,0.95,0.92}

\lstdefinestyle{mystyle}{
    backgroundcolor=\color{backcolour},   
    commentstyle=\color{codegreen},
    keywordstyle=\color{magenta},
    numberstyle=\tiny\color{codegray},
    stringstyle=\color{codepurple},
    basicstyle=\ttfamily\footnotesize,
    breakatwhitespace=false,         
    breaklines=true,                 
    captionpos=b,                    
    keepspaces=true,                 
    numbers=left,                    
    numbersep=5pt,                  
    showspaces=false,                
    showstringspaces=false,
    showtabs=false,                  
    tabsize=2
}
 
\usepackage{balance}

\begin{document}

\title{In Situ Network and Application Performance Measurement on Android Devices and the Imperfections}

\author{Mohammad A. Hoque}
\affiliation{
	\institution{University of Helsinki, Finland}
}
\email{mohammad.a.hoque@helsinki.fi}

\author{Ashwin Rao}
\affiliation{
	\institution{University of Helsinki, Finland}
}
\email{ashwin.rao@helsinki.fi}
\author{Sasu Tarkoma}
\affiliation{
	\institution{University of Helsinki, Finland}
}
\email{sasu.tarkoma@helsinki.fi}


%
\begin{abstract}
Understanding network and application performance are essential for debugging, improving user experience, and performance comparison. Meanwhile, modern mobile systems are optimized for energy-efficient computation and communications that may limit the performance of network and applications. In recent years, several tools have emerged that analyze network performance of mobile applications in~situ with the help of the VPN service. There is a limited understanding of how these measurement tools and system optimizations affect the network and application performance. 
In this study, we first demonstrate that mobile systems employ energy-aware system hardware tuning, which affects application performance and network throughput. We next show that the VPN-based application performance measurement tools,  such as Lumen, PrivacyGuard, and Video Optimizer, aid in ambiguous network performance measurements and degrade the application performance. Our findings suggest that sound application and network performance measurement on Android devices requires a good understanding of the device, networks, measurement tools, and applications. 
\end{abstract}

\settopmatter{printacmref=false}

\maketitle

\input{edgesys/introduction.tex}

\input{edgesys/systemopt.tex}
\input{edgesys/impact.tex}

\input{edgesys/discussion.tex}

\vspace{-2mm}
\bibliographystyle{plain}
\bibliography{references}


\input{edgesys/appendix.tex}
\end{document}

%% file: edgesys/introduction.tex
\section{Introduction}
\label{sec:intro}

In~situ Internet traffic measurement tools,   such as Video Optimizer (VoP)~\cite{Qian:2011:PRU},  Lumen~\cite{Razaghpanah:2017}, PrivacyGuard (PvG)~\cite{Song:2015:PVP}, and MopEye \cite{Wu:2017:MOM:3154690}, are essential for debugging, improving user experience, and performance comparison of mobile applications. The alternative is rooting the device and using \emph{tcpdump} for offline analysis.

The above traffic measurement tools shed light on the network and application performance. However, they may also contribute to imperfect and ambiguous results, as we might measure something which we do not intend to measure. Studying the sources of these imperfections is vital to calibrate the measurement procedures and to improve the tools. At present, there is a limited understanding of the impact of in~situ mobile Internet traffic measurement tools and how device hardware optimization affects the network and application performance.

In this work, we quantify the performance impact of system hardware optimization and also evaluate the impact of VoP, Lumen, and PvG on network performance metrics, and application traffic. We focus on these three applications, as they exemplify state-of-the-art traffic measurement and analysis tools. These tools have similar designs and use the Android VPN interface. However, they do not route the traffic to a remote VPN server. 
VoP~\cite{vop}, formerly known as ARO~\cite{Qian:2011:PRU}, is a popular open-source tool for collecting traffic from mobile devices without rooting the device, and it also enables various diagnosis and optimization of applications, network, CPU and GPU~\cite{vop} through offline analysis. 
In contrast, Lumen and PvG are two online traffic analysis tool helping users to find privacy leaking incidents. Lumen also provides insights on the TLS usage of mobile applications~\cite{Razaghpanah:2017}, the CDN usage by mobile applications~\cite{8485872}, and the  DNS~\cite{Almeida2017DissectingDS}. MopEye is another similar application. It is currently unavailable in the Google Play Store and also in popular source code hosting websites, such as GitHub.

This article investigates the imperfections in traffic measurements on Android devices due to system optimization and in~situ traffic measurement tools. We demonstrate that sound Internet traffic measurement requires a thorough understanding of the device, tools, and applications. Note that we do not aim to establish whether a particular tool is the best or worst. Our key observations are as follows. 

\emph{(1)} Mobile systems employ CPU and WiFi transmit power optimization triggered by the battery level.
We observe that the CPU optimization techniques, such as CPU hot-plugging and dynamic frequency scaling, mostly affect network I/O, while WiFi optimization, i.e., dynamic modulation scheme, affects the uplink throughput. These optimizations deteriorate application performance and network throughput. Charging the device, when the battery level is below 20\%, does not improve the network performance. Therefore, one must be aware of the adaptive performance characteristics of mobile devices while conducting experiments (Section \ref{sec:systemopt}).

\emph{(2)} Although it is expected that VPN-based tools would provide degraded network performance as the packets spend more time on the device \cite{Qian:2011:PRU,Razaghpanah:2017,Song:2015:PVP},  we may estimate ambiguous latency and throughput in the presence of the VPN-based tools. For example, in the presence of PvG, SpeedCheck~\cite{speedcheck} estimates on-device latency instead of the network latency. Similarly, VoP doubles the uplink throughput estimates. The sources of these ambiguities are the implementation of the measurement tools, as we present in Section \ref{sec:impact}. VoP also delays the outgoing traffic, and PvG delays the incoming traffic. Therefore, to avoid such pitfalls in network and application performance measurements, one must have a good understanding of these applications and tools.

\emph{(3)} Furthermore, all these VPN-based applications fail to apply the application intended optimization through socket options and thus affect the application performance, as we demonstrate for the outgoing TCP traffic in Section~\ref{sec:impact}.

Finally, we summarize the sources of the above ambiguous or imperfect measurement results (Section \ref{sec:inefficiency}).

\begin{figure}[t]
\centering
  \includegraphics[width=1.0\linewidth]{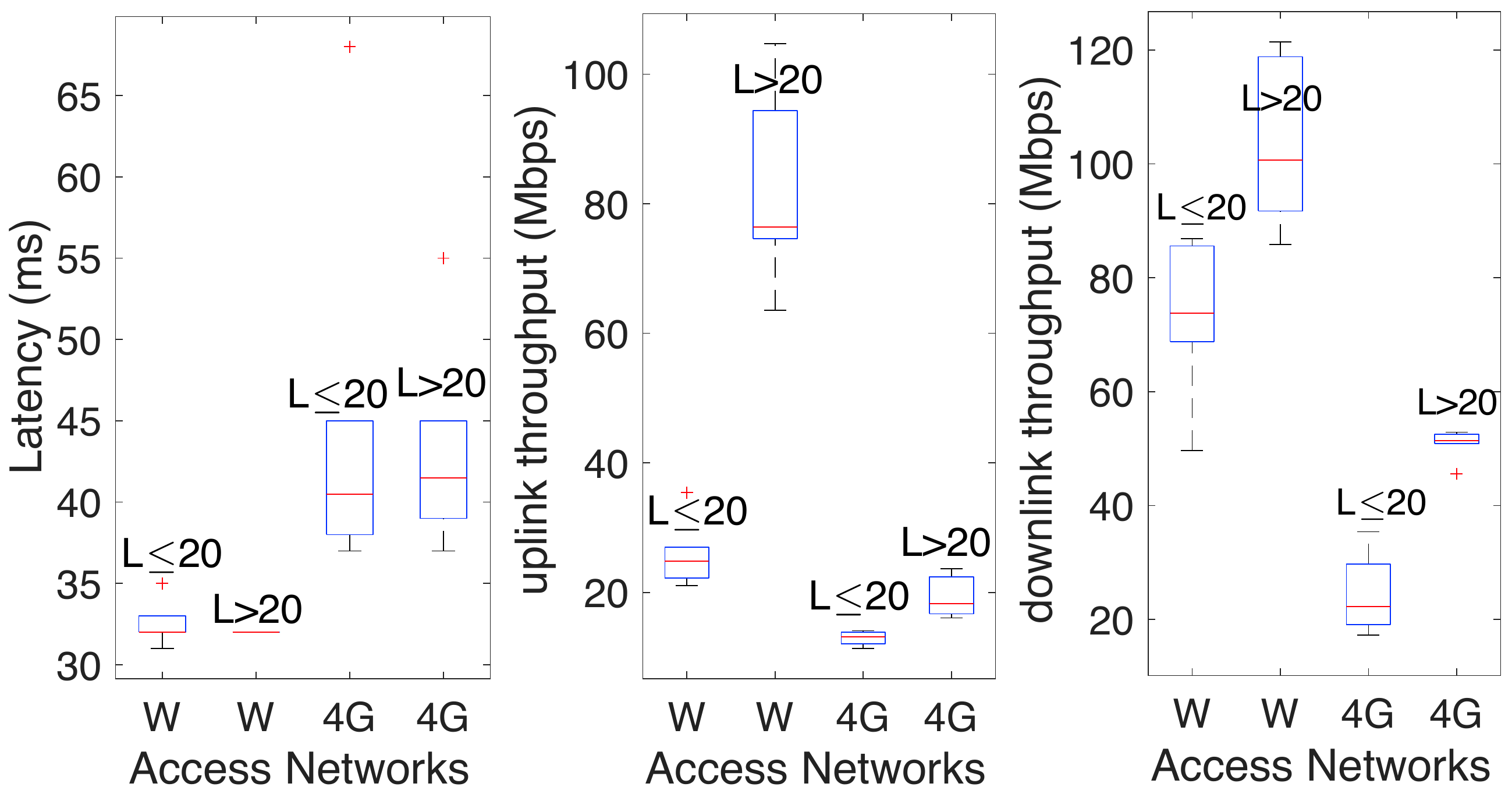}
  \vspace{-3mm}
\caption{{\bf Impact of battery level.} {\sl We consider two battery level (L) ranges, L$\le$20\% \& L$>$20\%, on Nexus 6 over WiFi (W) and LTE (4G).}}
\label{fig:optimization}
\end{figure}

%% file: edgesys/systemopt.tex
\section{Impact of System Optimization}
\label{sec:systemopt}

Android devices may come with advanced CPU governors that save energy by hot plugging and unplugging of CPU cores, as supported by modern Linux kernels~\cite{cpuhotplug}.
Apart from workload characteristics, the devices may also consider the status of the battery to employ the CPU cores. We look into the impact of such off-the-shelf system optimization on network latency and throughput on Nexus 6. 

During our measurements with Nexus 6, we have found that two of the four cores remain offline when the battery discharges to below 20\%, and the active cores operate at the maximum frequency of 1.73~GHz. 
When the battery level is above 20\%, all the four cores become active, and their maximum operating frequency increases to 2.65~GHz. Therefore, the battery level also prompts dynamic frequency scaling. 

We performed the following measurements to quantify the impact of this optimization on the network traffic characteristics.
Specifically, we used SpeedCheck~\cite{speedcheck} (paid) and measured the latency and throughput on Nexus~6 (Android 7.0) when the battery levels were above 20\% and below 20\%.
We performed the measurements using both WiFi and LTE.
Each of the above four scenarios was repeated ten times, and the results are presented in \autoref{fig:optimization}. \autoref{fig:optimization} shows that while hot unplugging of CPU cores on Android has a negligible impact on the latency, its impacts on throughput is significant. The availability of additional CPU cores, when the battery level is above 20\%,  improves the I/O performance across the two access technologies, WiFi and LTE.
Furthermore, WiFi uplink throughput improves almost four times when the battery level is above 20\% compared to when it is below 20\%.
The closer inspections of the MAC layer frames revealed that WiFi radio of the Nexus 6 switches from \emph{802.11ac} to \emph{802.11g} mode when the battery level drops below 20\%.  These performance limiting optimizations also affected the device responsiveness for various applications, such as browsing and streaming.  This also implies that modern Android devices adapt the physical layer mechanisms similar to the iOS devices\footnote{https://www.forbes.com/sites/ewanspence/2017/12/20/apple-iphone-kill-switch-ios-degrade-cripple-performance-battery/} to avoid unexpected shutdown of the devices \cite{8720247} and to improve battery life.

\begin{figure}[t]
\centering
  \includegraphics[width=1.0\linewidth]{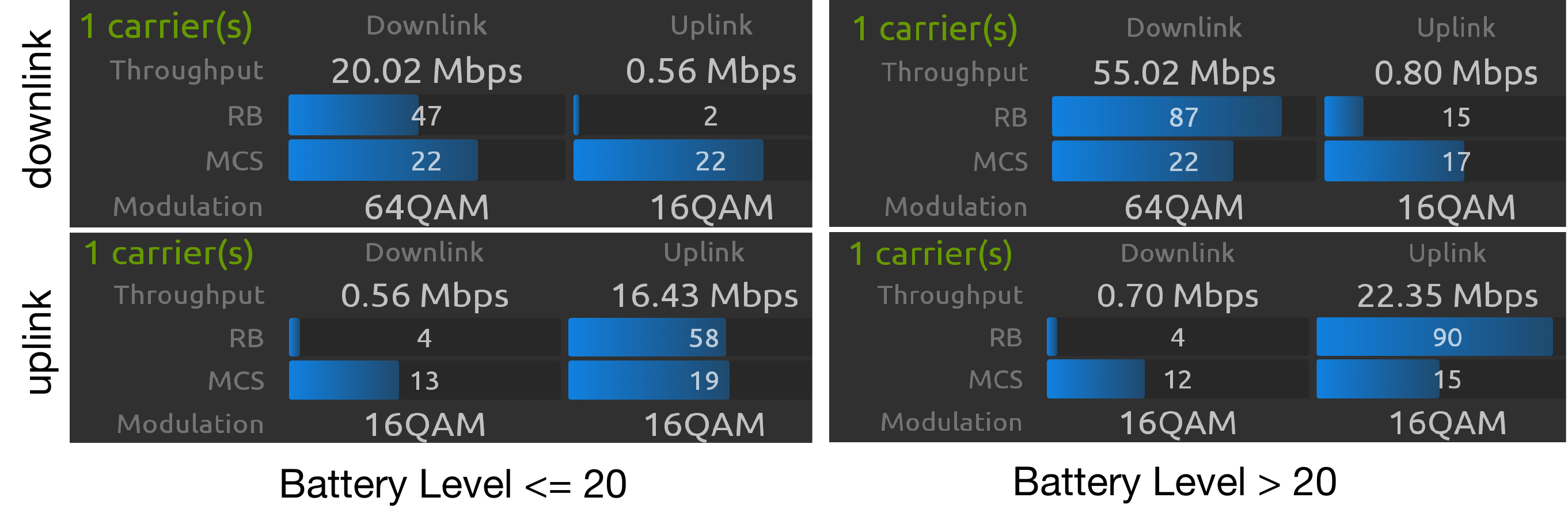}
\caption{{\bf Impact of battery level on LTE modulation scheme. These snapshots are from a single uplink and downlink throughput measurement.}}
\label{fig:ltemod}
\end{figure}

Figure \ref{fig:optimization} also depicts that the downloading speed of SpeedCheck over LTE doubles when the battery level is higher than 20\%. Similar to WiFi, we further looked into the physical layer modulation scheme used by the mobile device in the LTE network. We rooted Nexus 6 and installed Network Signal Guru \cite{netsiguru} that samples LTE physical layer parameters after every 500 ms.  \autoref{fig:ltemod} shows that the modulation schemes were always 16QAM (Quadrature Amplitude Modulation) and 64QAM for uplink and downlink, respectively, during the throughput measurements. The other attributes in the figure are discussed in \autoref{sec:mvs}. Nexus 6 employs three optimization techniques,   triggered by the battery level, which affect the network and application performance.  Charging the device, when the battery level is below 20\%, does not improve the throughput either on WiFi or LTE and application performance. The optimization may vary from device to device.


%% file: edgesys/impact.tex
\section{Impact of Measurement Tools}
\label{sec:impact}

\begin{figure}[t]
    \centering
    \includegraphics[width=0.80\linewidth]{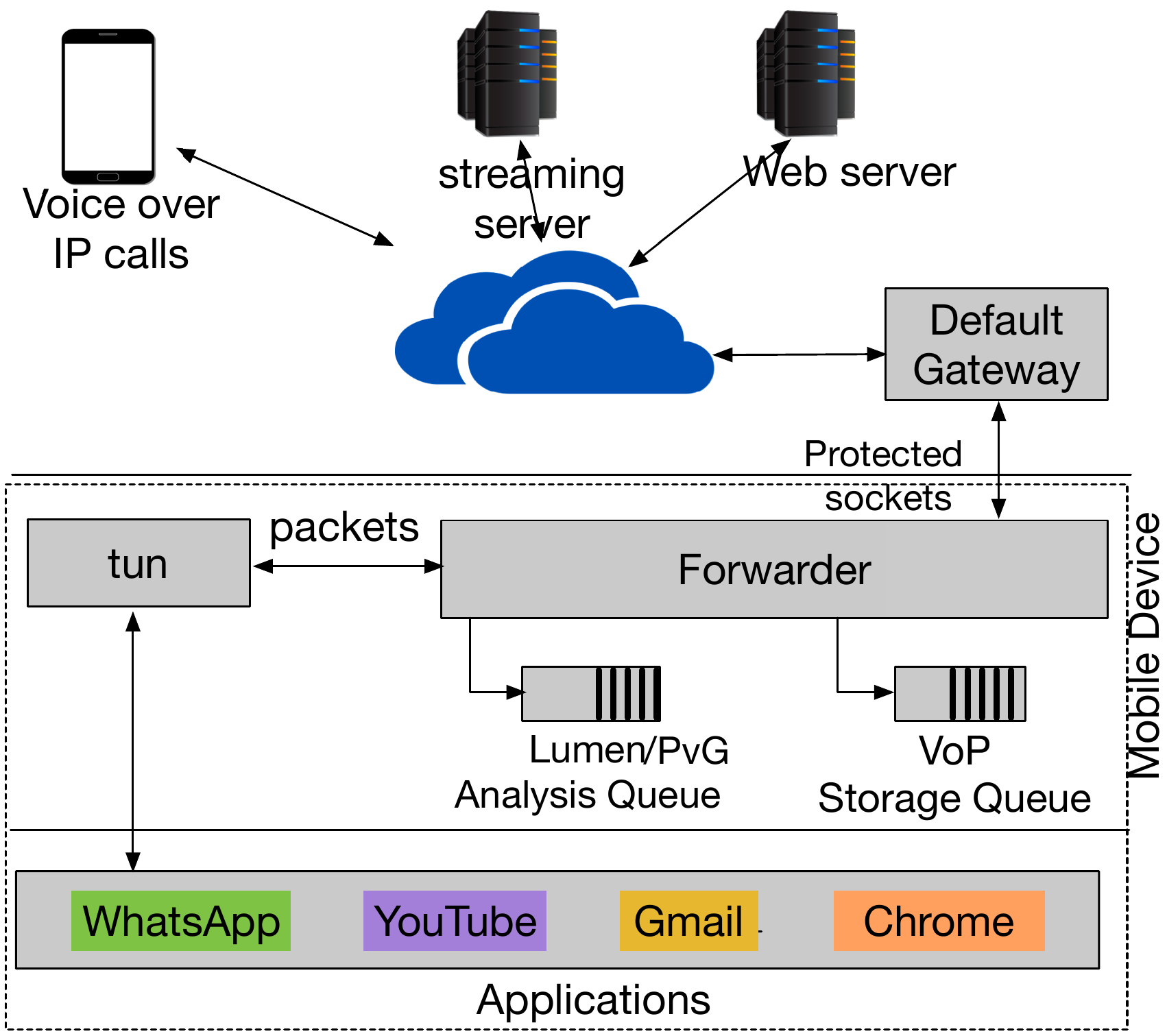}
    \caption{{\bf The system components of VoP, Lumen, and PvG   for Android.}{\sl The newly created sockets are protected so that the Forwarder generated packets are not in a loop.}}
    \label{fig:lumenVoP}
\end{figure}

\begin{figure*}[t]
\centering
  \includegraphics[width=0.8\linewidth]{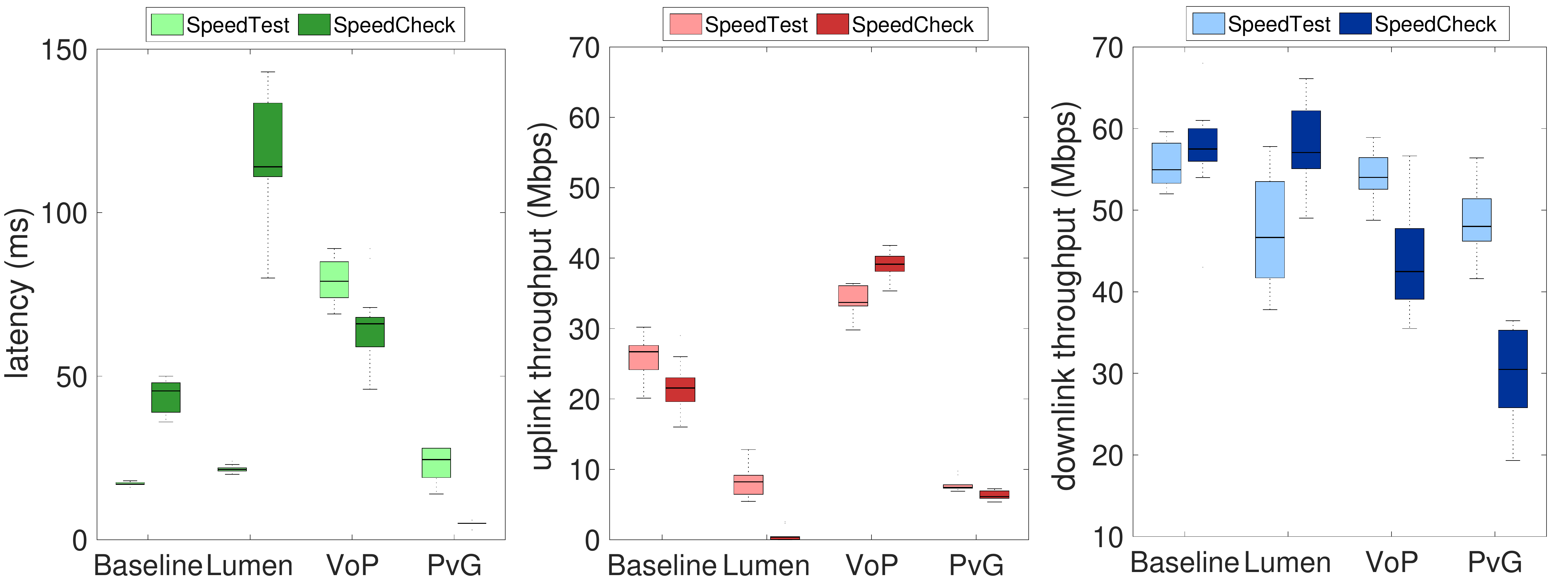}
  \vspace{-4mm}
\caption{{\bf Impact on LTE network latency and throughput.} {\sl We used SpeedCheck and SpeedTest on Nexus 6 in the presence of Lumen (Lum.), VoP, PvG, and Baseline, i.e., without any localhost VPN.}}
\label{fig:netperf}
\end{figure*}

\subsection{In-situ Traffic Measurement Tools}
\label{sec:measurementtools}

The forwarder and the packet inspector are two components of the VPN-based in~situ traffic measurement tools exemplified by VoP,  Lumen, and PvG, as shown in \autoref{fig:lumenVoP}.

The primary role of the forwarder is to forward (i) the packets received from Android applications to the Internet, and (ii) the packets received from the Internet to the Android applications.  The forwarder also copies those packets to the inspection queue to isolate traffic analysis from the path of the packet. 

The forwarder essentially creates a new TCP socket on seeing a TCP SYN packet from the VPN interface. The forwarder in Lumen and VoP establish a socket connection with the remote server using \texttt{connect()} API before sending SYN-ACK to the application. PvG, on the other hand, establishes socket connection after replying with SYN-ACK. Later, we demonstrate how these implementations affect network performance measurements. The forwarder creates a new UDP socket when it detects a new UDP flow.
These newly created sockets are protected so that packets from the newly created flows do not loop the \emph{tun} interface \cite{vpnprot}. 

A packet inspector is responsible for inspecting the packets in its queue. 
In the case of Lumen and PvG, the packet inspector performs the privacy analysis on the packets, whereas the VoP's inspector sends packets to the desktop application.



In the later sections, we quantify the impact of VoP, Lumen, and  PvG on (a) the network performance, and (b) the network characteristics of applications.

\subsection{Addressing Biases}

We took the following steps to ensure that the measurement results presented in the upcoming sections are not the artifacts of misconfigured tools and the measurement setup.

{\sl (i) Battery level.}
For the upcoming measurements, we ensured that the devices had more than 80\% charge. This is because mobile devices might restrict resources based on the battery level, as we have shown in \autoref{sec:systemopt}.

{\sl (ii) Throughput throttling.}
VoP also offers to throttle downlink and uplink traffic. All the measurements in this paper were conducted without any throughput throttling. 


{\sl (iii) Software Auto Update.}
During the experiments, application and the auto system updates were disabled on mobile devices.

{\sl (iv) Advertisements.}
 We have purchased without ad subscriptions of SpeedCheck and SpeedTest to avoid any biases caused by the free versions. 

\begin{figure*}[!h]
 \subfigure[Baseline]{\label{fig:voipbase}\includegraphics[width=0.3\linewidth]{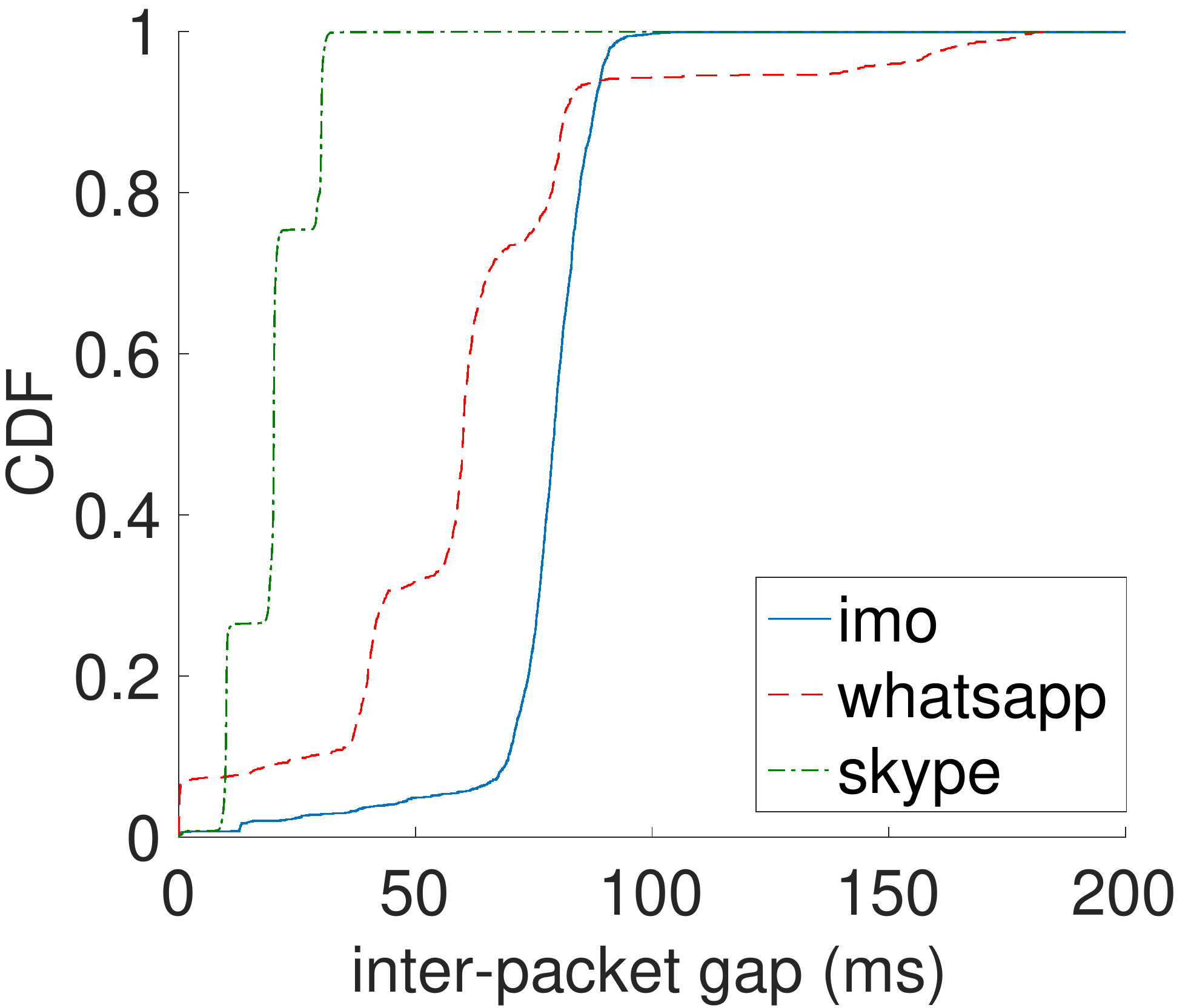}}
 \subfigure[VoP]{\label{fig:voipvop}\includegraphics[width=0.28\linewidth]{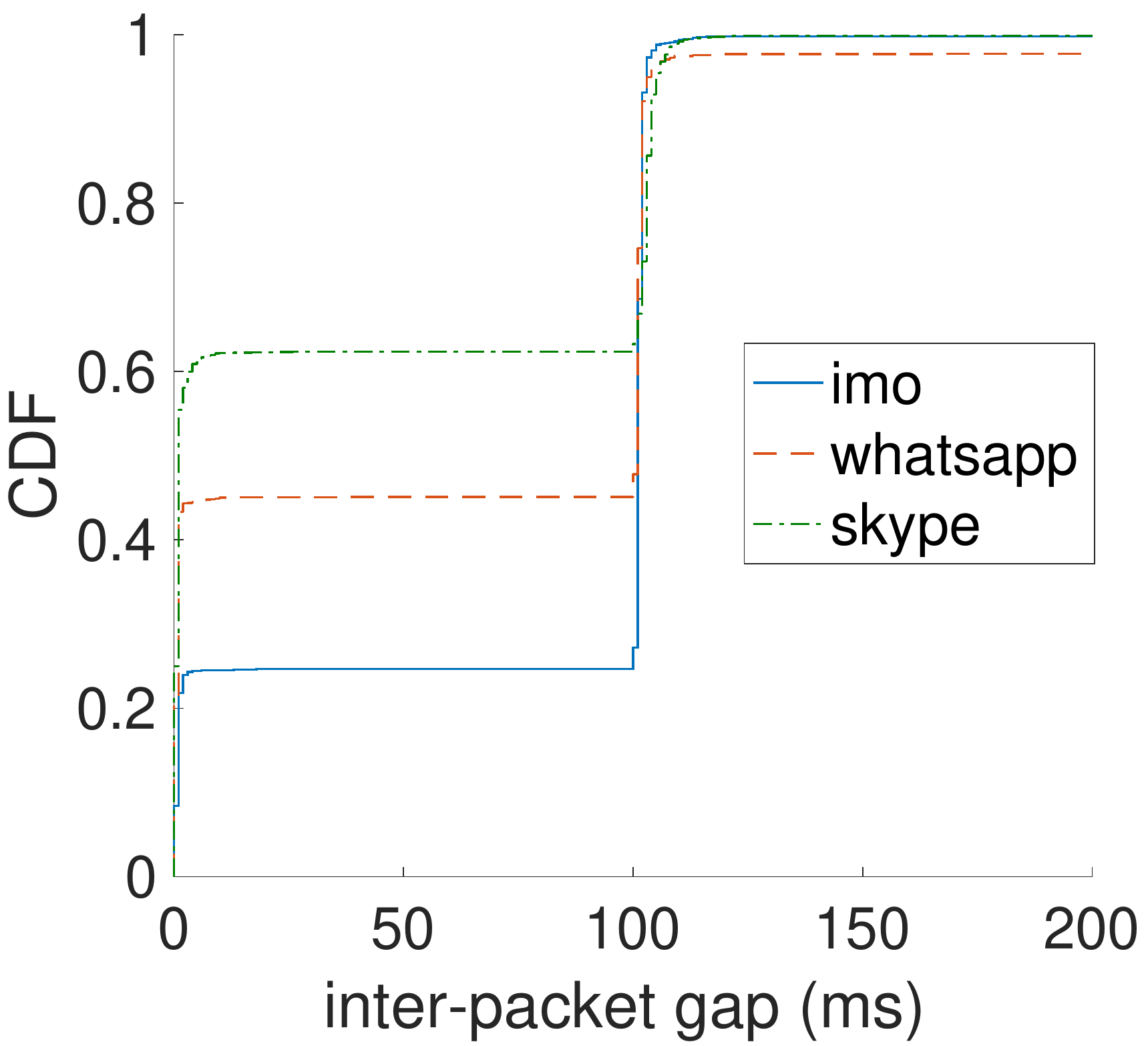}}
 \subfigure[Lumen]{\label{fig:voiplumen}\includegraphics[width=0.28\linewidth]{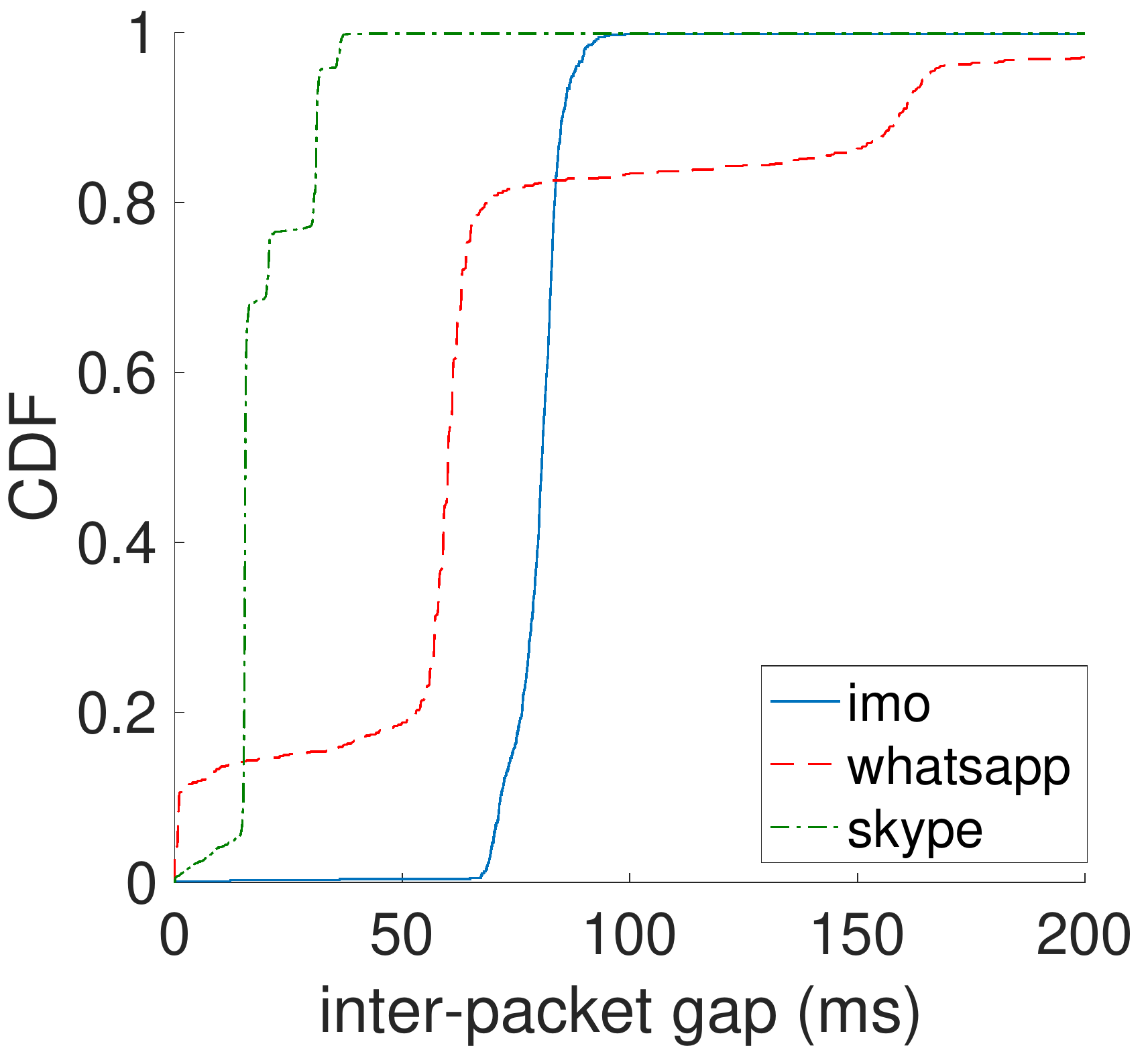}}
\caption{{\bf Inter-packets gaps of the VoIP applications}. {\sl Baseline refers to the  measurements without any localhost VPN.}}
\label{fig:ipVoiceIP}
\end{figure*}

 \subsection{Impact on Network Performance}
\label{sub:latban}


 This section explores the network performance using   SpeedCheck \cite{speedcheck} and SpeedTest \cite{speedtest}. These two applications work as the traffic load generator without any VPN-based tools and in the presence of the listed VPN applications. Without any VPN scenario gives the baseline performance. SpeedCheck connects to its servers in Germany, and SpeedTest connects to the severs in the LTE operator network within a few kilometers from the mobile device. The measurements were repeated ten times.

\emph{(1) Latency.} \autoref{fig:netperf} (left) compares the network latency reported by two applications in the presence of the VPN-based tools. From the \emph{tcpdump} traces, we have identified that SpeedTest uses 10-12 requests/responses of few bytes (less than 100 Bytes) over a TCP connection to estimate the latency. SpeedTest estimates the baseline latency of 16-18 ms. This is expected, as the server was located at the operator's network. It experiences 3-5 ms additional latency in the presence of Lumen and PvG, whereas VoP increases the latency by three-fold. This is due to the energy optimization strategy adopted by VoP, which we discuss in the upcoming sections.

In contrast, SpeedCheck reports the median baseline network latency of about 45~ms. From the corresponding \emph{tcpdump} traces, we have identified 10 empty and consecutive TCP flows (without any data exchange) for each latency measurements. These flows suggest that SpeedCheck uses TCP \texttt{connect()} API to measure the latency. Both VoP and Lumen increase the median latency significantly. We speculate that these two take more time to set up new TCP flows. However, SpeedCheck underestimates the latency in the presence of PvG, which is the consequence of the sending SYN-ACK by the PvG forwarder before the connection is established with the remote server, as discussed in Section \ref{sec:measurementtools}.


\emph{(2) Uplink Throughput.} Figure \ref{fig:netperf} (center) depicts that SpeedTest estimates higher uplink baseline throughput, as the server is in the LTE operator network. It uses multiple parallel TCP connections to estimate the throughput. Both Lumen and PvG reduce the throughput of  SpeedTest/SpeedCheck by half compared to the baseline measurements. However, Lumen severely affects the uplink throughput measurements of the SpeedCheck. It uses a single TCP connection and sends a large amount of data. From an exception in the debug log, we characterized that Lumen's forwarder cannot handle such volume. Interestingly, VoP doubles the uplink throughput of both applications. 

\emph{(3) Downlink Throughput.} Figure \ref{fig:netperf} (right) demonstrates that SpeedTest measures similar downlink throughput in the presence of the VPN tools to the baseline. Lumen aids the highest throughput measurements with SpeedCheck. However, VoP and PvG degrade the throughput of SpeedCheck significantly.

The typical network measurement tools, such as SpeedCheck and SpeedTest, can have different methods to estimate the latency and throughput. While their baseline estimates are reasonable, their estimates vary according to the implementation of the VPN tools. 


\vspace{-2mm}
\subsection{Impact on Realtime Application (UDP)}
\label{sub:netcharacudp}

In this section, we investigate the traffic from three realtime applications;  IMO, WhatsApp, and Skype. The versions of the apps used are presented in \autoref{tab:datarates}.
While these applications fall into the broad category of messaging applications, their varying traffic characteristics help us to study the impact of the design of VoP and Lumen. We could not use these applications in the presence of PvG in several trials. 
We used a rooted Nexus 6 (Android 7.0) and a non-rooted LG G5 (Android 8.0) for these measurements.



These apps exchange bi-directional encrypted UDP traffic. The conversations were two minutes long over LTE, and we ran 3 iterations in each of the following scenarios. We investigate their inter-packet gaps and bitrates. 
As the baseline, we initiated conversations between Nexus 6 and LG G5 using these apps without VoP or Lumen and captured traffic using \emph{tcpdump} on Nexus 6. We then repeated the experiments with VoP running on Nexus 6 and collected traffic from VoP. Finally, we used Lumen. Since Lumen does not store traffic, we captured traffic with \emph{tcpdump} on Nexus 6.

\begin{table}[t]
 \centering
 {
\begin{tabular}{ |r | c | c | c |}
\hline
                  & {\bf Baseline} & {\bf VoP} & {\bf Lumen} \\
{\bf Application} & (in/out)  & (in/out)  & (in/out)  \\
\hline
WhatsApp (v2.18) & 21/24~kbps & 23/16~kbps & 20/22~kbps   \\
IMO (v9.8)       & 14/15~kbps & 14/13~kbps & 13/14~kbps\\
Skype (v8.41)    & 60/50~kbps & 55/44~kbps & 48/44~kbps\\
\hline
\end{tabular}}
\addtolength{\tabcolsep}{2pt}
\caption{Average bitrates of UDP traffic flows from VoIP applications.}
\label{tab:datarates}
\end{table}

\emph{Baseline Results.}
Figure \autoref{fig:voipbase} shows that IMO has the highest inter-packet gaps, and Skype packets have the smallest gaps. These apps also have distinct data rates with Skype having the highest data rate, as shown in \autoref{tab:datarates}. 

\emph{Impact of VoP.}
Compared to the baseline packet-gaps in Figure \autoref{fig:voipbase}, VoP significantly alters the inter-packet gaps of outgoing UDP packets, as shown in Figure \autoref{fig:voipvop}. Most of the outgoing packets across all applications have an inter-packet gap of about 100~ms. In contrast, the incoming packets have had similar distributions to the baseline. This delay is similar to the latency measurements with VoP discussed earlier. 
\autoref{tab:datarates} shows that the outgoing data rates of Skype and Whatsapp reduce significantly, which we speculate to be a consequence of the delays introduced by VoP. 

\emph{Impact of Lumen.}
Figure \ref{fig:voiplumen} shows that with Lumen the inter-packet gaps of the outgoing packets are similar to the baseline measurements. Besides, the applications experience similar bitrates to the baseline and when using Lumen as shown in  \autoref{tab:datarates}.

\subsection{Impact on Realtime Application (TCP)}
\label{sub:netchartcp}
We used Periscope (v1.24) to study the impact of VoP and Lumen on realtime TCP flows. Periscope's live broadcast did not work in the presence of PvG. Periscope broadcasts over LTE across three different scenarios.  We capture traffic on Nexus 6 using \emph{tcpdump} for baseline and Lumen scenarios.

\begin{figure}
  \includegraphics[width=0.49\linewidth]{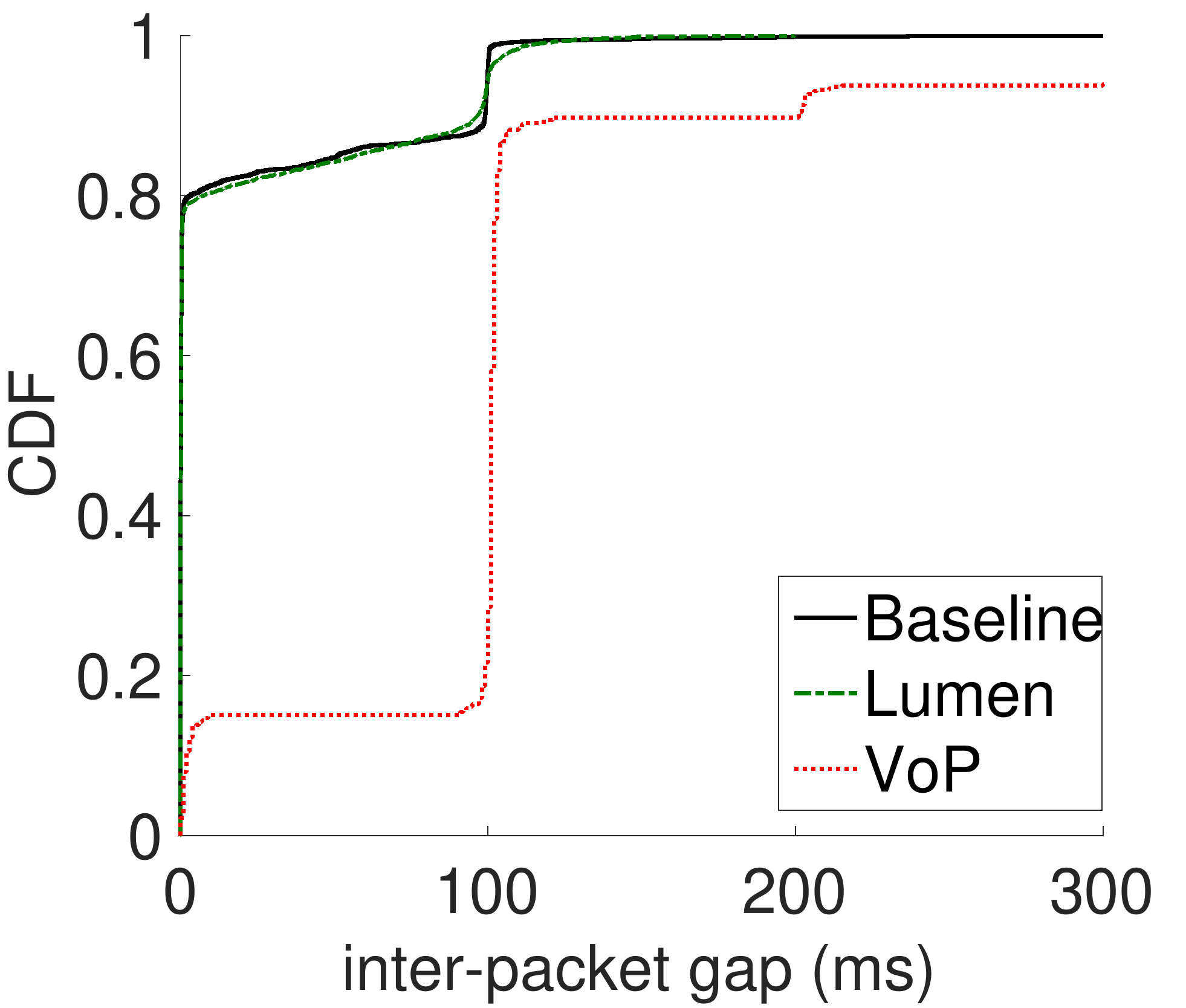}
  \includegraphics[width=0.49\linewidth]{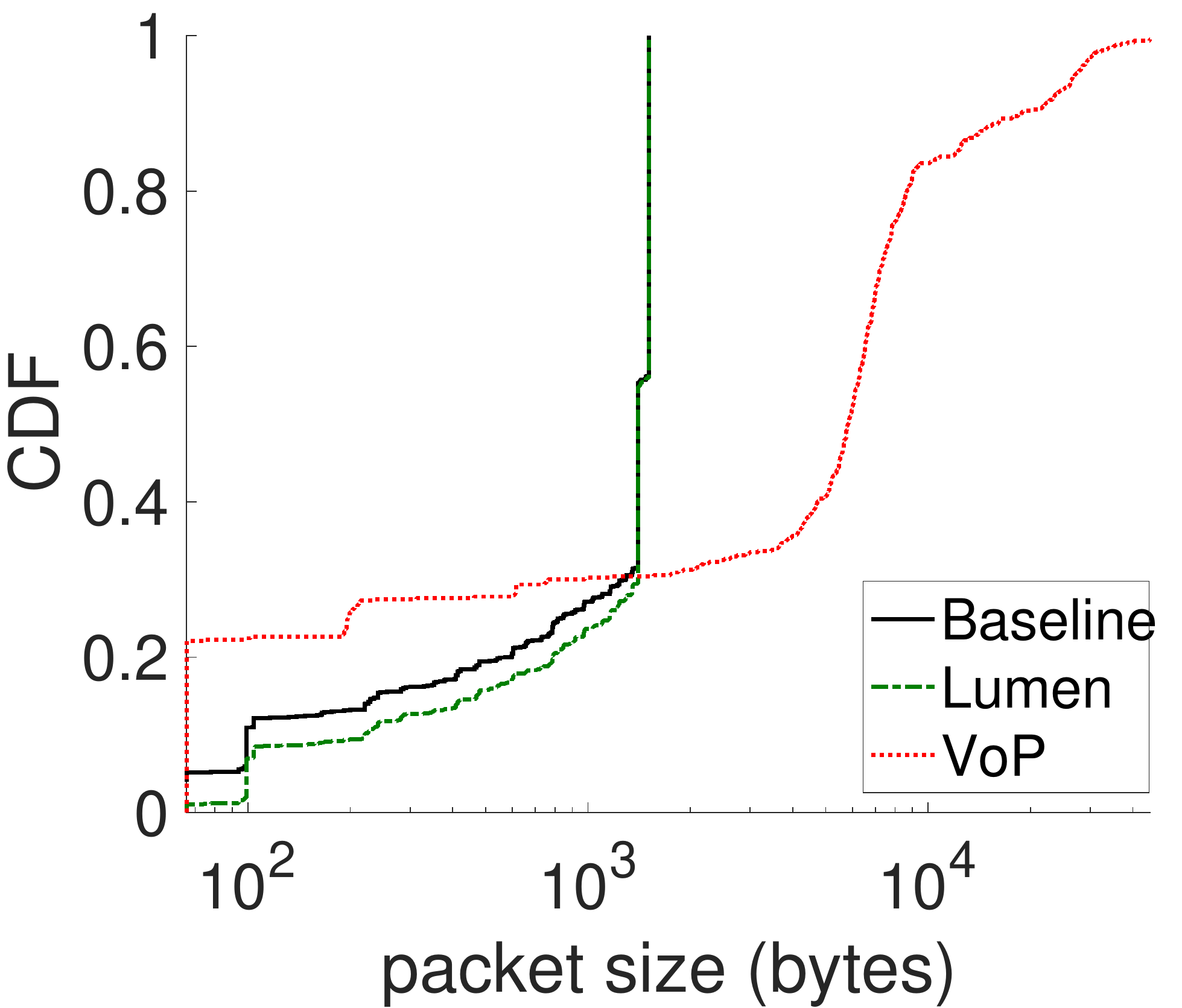}
\caption{{\bf Properties of  uplink Periscope TCP flows.}}
\label{fig:periscopes}
\end{figure}

Similar to our observations for UDP traffic, we observed 100~ms inter-packet gap, as shown in \autoref{fig:periscopes} (left).
From the distribution of packet size in \autoref{fig:periscopes} (right) (collected by VoP), we notice that more than 70\% packets captured by VoP are larger than 1500 bytes. From Traffic traces, we have identified that VoP creates packets of a maximum of 65549 bytes for Periscope, and the uplink throughput measurements flow from SpeedCheck.

 From the source code in Github, we have identified that VoP forwarder implements the maximum segment of 65535 bytes for the TCP flows. It accumulates traffic from the client application, and the segments reach the maximum size very quickly with very high bitrate traffic. This also explains how VoP aids in higher uplink throughput measurements presented in Section \ref{sub:latban}. Nevertheless, these massive TCP segments are eventually fragmented once written to the socket. Lumen has a very negligible impact on packets.

%% file: edgesys/discussion.tex
\subsection{Analysis with Socket Options}
\label{sec:sockopt}


In this section, we investigate the performance of the VPN-based tools in processing the flows with TCP\_NODELAY (Nagel's algorithm) socket option on Nexus 6. We specifically look into this option, as it has a direct impact on the local delay and thus affects the performance of web browsing and other realtime applications, such as live broadcasting, crypto/stock exchange applications, on mobile devices. We developed a separate traffic generating application that creates two blocking TCP sockets enabled and disabled  Nagle's algorithm.  The application sends 1300 bytes data over LTE after every 20 ms to a remote server at the university campus. The application also receives data from the remote server after every 20 ms in separate TCP sessions. 

\begin{figure}[t]
\centering
  \includegraphics[width=1.0\linewidth]{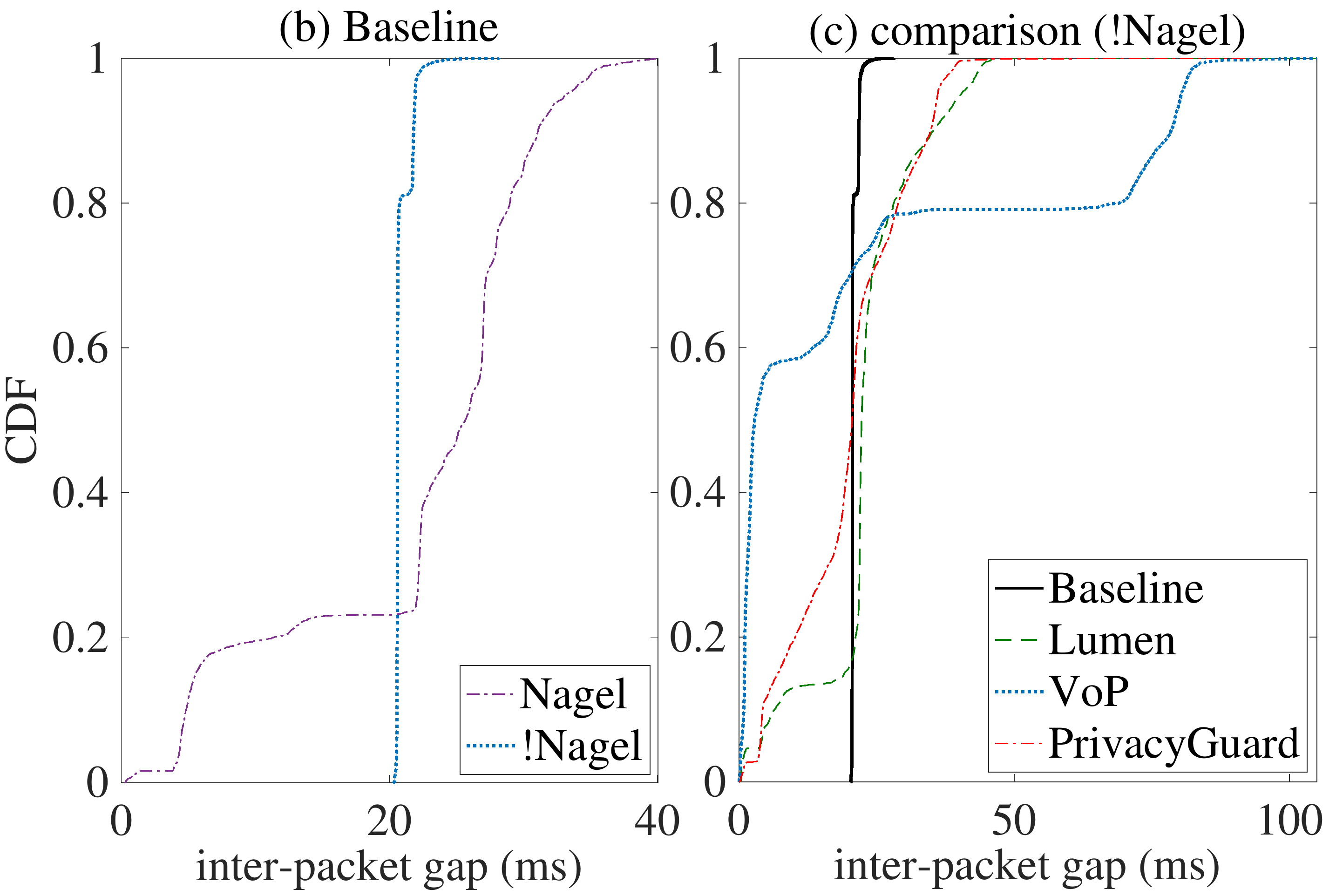}
\caption{{\bf Distributions of the outgoing packet gaps observed at the network interface.}}
\label{fig:vpn_test_outgoing}
\end{figure}

\begin{figure}[!h]
\centering
\includegraphics[width=1.0\linewidth]{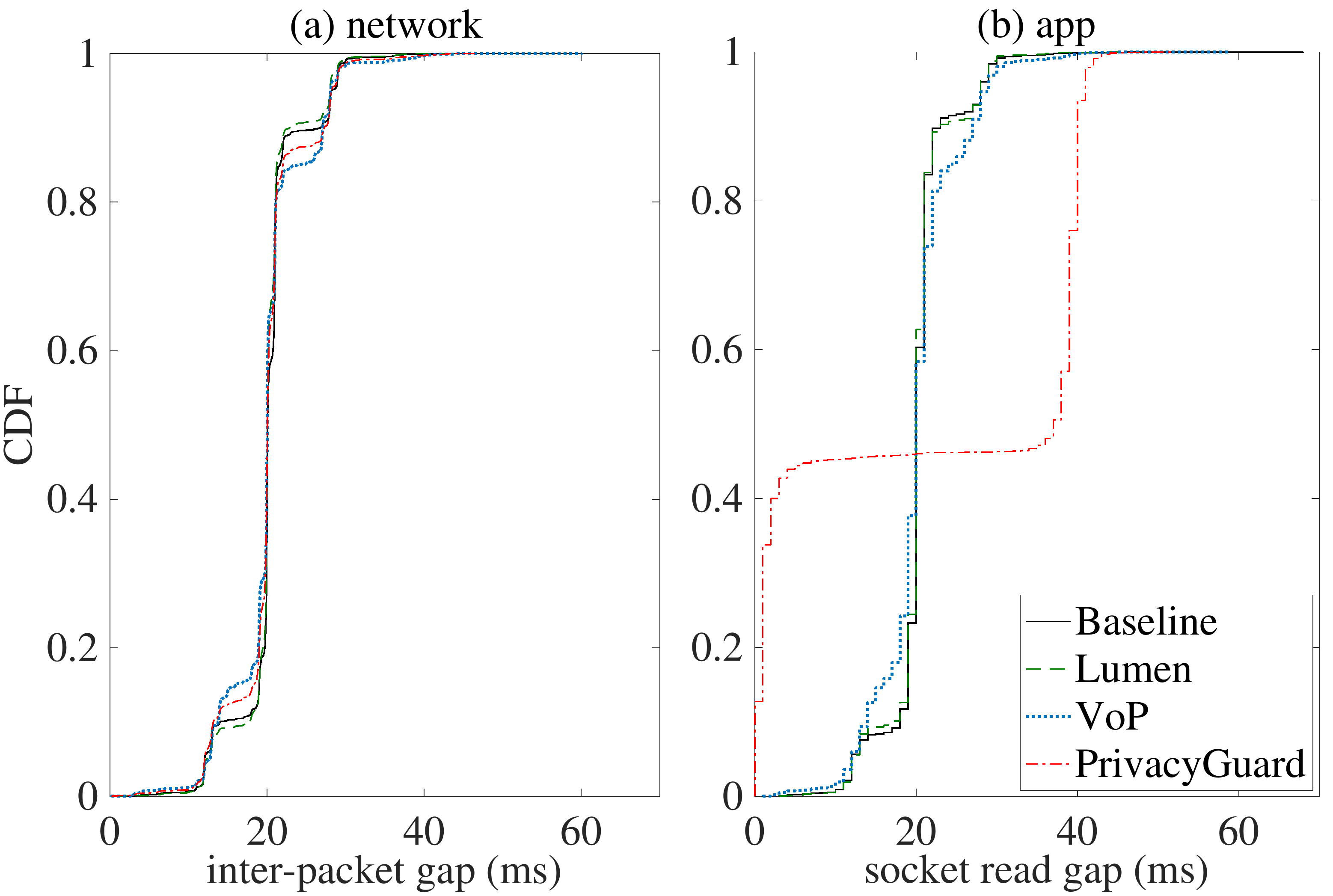}
\caption{{\bf Distributions of incoming packet gaps observed at the network interface and application.}}
\label{fig:vpn_test_incoming}
\end{figure}

\emph{Performance of VPN-based Tools.} 
Figure \ref{fig:vpn_test_outgoing}(a) compares the outgoing inter-packet gap of the application flows; having Nagel's algorithm enabled and disabled. When Nagel's algorithm is enabled, more than 70\% of the packets sent from the application have more than 20 ms delays at the network layer. In the presence of VPN applications, disabling Nagel's algorithm by the application does not improve the delay compared to the baseline (Figure \ref{fig:vpn_test_outgoing}(b)).   Interestingly, VoP's packet gap reduces, as it receives packets from the local TCP/IP stack without delay. From traffic traces, we have identified that these VPN-based tools do not disable Nagel's algorithm while establishing socket connections.

Figure~\ref{fig:vpn_test_incoming} shows the performance of the VPN applications for incoming traffic. The application receives data at almost similar gaps observed at the network interface. However, in the presence of PvG, the application receives 40\% packets at late. The packet-gaps patterns suggest that it uses a fixed interval to read the VPN interface similar to VoP. 

The investigations in this section reveal that the VPN-based tools do not set the TCP/IP socket options as intended by the other user applications. Consequently, they can misguide the developers and degrade application performance. For example, SpeedTest disables Nagel's algorithm or sets the TCP\_NODELAY socket option to send tiny packets to measure the network latency. Findings in this section explain the higher latency experienced by SpeedTest in Section \ref{sub:latban}.

\vspace{-1mm}
\section{Sources of Imperfection}
\label{sec:inefficiency}

Mobile system optimizations affect downlink and uplink throughput, whereas the VPN-based tools mostly affect the uplink throughput and latency, i.e., they mostly affect the outgoing traffic. In this section, we summarize the sources of such measurement results. 

\emph{Energy-Aware Optimization.} Energy-aware system optimization can affect the network performance by limiting the network I/O and by applying adaptive modulation schemes. Therefore, it is wise to perform such measurements when the battery is fully charged. VoP, Lumen, and PvG rely on different sleeping techniques to optimize their energy usage.
The additional latency introduced by VoP on outgoing packets is the artifact of using a fixed sleep interval of 100 ms in the main VPN thread. This delay further contributes to large outgoing packets for higher bitrate uplink traffic and energy consumption for fragmentation. PvG also introduces a fixed delay for the incoming traffic.   
Regardless, these delays affect not only the quality of the measurements but also the quality of experience when using other user applications. 

\emph{Forwarder.}
In situ VPN-based measurement tools are middleboxes that tap the packets using the VPN interface. These applications, therefore, implement a forwarder which primarily consists of three threads: the main VPN thread, and two-socket reader/writer threads.
The reader/writer threads continuously iterate through a list of live sockets, which contributes to the delays. The forwarder also implements a flow state machine for each flow and constructs/de-constructs the packets. The implementation of the forwarder affects the latency and throughput measurements. We have also shown that the characteristics of the newly created flows and their packet headers might not be the same as those generated by the applications. The reason is that the socket options must be set before the connection establishment.  

\section{Conclusions}
\label{discussion}


In this preliminary work, we investigated the challenges in measuring network performance in the presence of system optimizations and state-of-the-art application performance measurement tools on Android devices. System optimizations limit the performance of the hardware components and thus the applications, which in turn result in confusing measurement results.  It can be argued that VoP is mostly for the developers, and therefore,  incurring higher delays should not a problem. Similarly, frequent massive content uploading is rare, and 3-4 ms additional latency is acceptable. Nevertheless, these imperfections can significantly affect the outcome of traffic measurement studies. An acceptable latency also depends on the application type. A user can benefit significantly from 1-millisecond latency improvement for the financial and other realtime applications.  
Therefore, there is still room for improvement in such tools. For instance, VoP and PvG can follow Lumen's adaptive sleeping algorithm for reducing the gaps in the outgoing and incoming packets, respectively. All of them can adopt some default socket options to mitigate the performance issues with the outgoing TCP traffic. Along with the measurement tools, it is necessary to understand the presence of various system optimization techniques which may affect network performance.

%% file: edgesys/appendix.tex


\section{LTE Radio Resource Allocation}
\label{sec:mvs}


In LTE networks, Physical Resource Block (RB) is considered as the unit of the radio resource. With 5 MHz bandwidth, there are 25 RBs. In an RB, there are 12 sub-carriers in the frequency domain. Each of the RBs can have either $7\times 12$ or $14\times12$ resource elements (REs), where 7 and 14 are the symbols, in the time domain, over 0.5 and 1 ms respectively using normal cyclic prefix (CP)~\cite{zyrenwhite}.

Now the amount of bits an RB can carry depends on the channel quality indicator (CQI) notification from the UE. Essentially, each CQI maps to a modulation and coding scheme according to Table~\ref{tab:cqi_index}. CQI indicates not only the channel quality but also a device's capability whether the device can receive data of a particular modulation and coding scheme or not.

The equations to compute the number bits an RB can hold for a certain CQI, and the number of RBs is required by an eNB to transmit a packet can be expressed as the followings.

\begin{equation}
    \label{eq:one}
    RB_{bits}= RE_{bits}\times n\times t_{s}\\
     = C_{CQI}\times M_{bits}\times n\times t_{s}
\end{equation}
In equation\ref{eq:one},  $M_{bits}$ is the bits for a modulation scheme, $n$ is the number of usable REs, and $t_s$ is the duration of time slot (0.5 or 1 ms). 
\begin{equation}
    \label{eq:two}
    RB_{n}= (PacketSize_{bits}+RLC_{bits}+MAC_{bits})/RE_{bits}
\end{equation}
\begin{table}[!h]
    \centering
    {\small
    \begin{tabular}{|c|c|c|c|}
    \hline
        CQI& Modulation & Real Bits ($N_m$) & $C_{CQI} = N/1024$ \\\hline

1    & QPSK    & 78&0.0762 \\\hline
2    & QPSK&    120&0.1171 \\\hline
3    & QPSK&193&    0.1884\\\hline
4    & QPSK&    308&0.3\\\hline
5    & QPSK&    449&0.4384\\\hline   
6    & QPSK    & 602 & 0.5879\\\hline
7    & 16QAM    & 378 & 0.3691\\\hline
8    &16QAM&    490&    0.4785\\\hline
9    &16QAM&616&0.6015\\\hline
10    &64QAM& 466    &    0.4550\\\hline
11    & 64QAM    & 567 & 0.5537\\\hline
12    &64QAM&    666&    0.6503\\\hline
13    & 64QAM&    772&0.7539\\\hline
14    &64QAM&    873&    0.8525\\\hline
15    &64QAM&    948&    0.9258\\\hline   \end{tabular}}
    \caption{Channel Quality Index (CQI), Modulation Scheme, and Coding Rate mapping ~\cite{kawsercqi}.}
    \label{tab:cqi_index}
\end{table}

\begin{figure}[]
    \centering
    \includegraphics[width=0.8\linewidth]{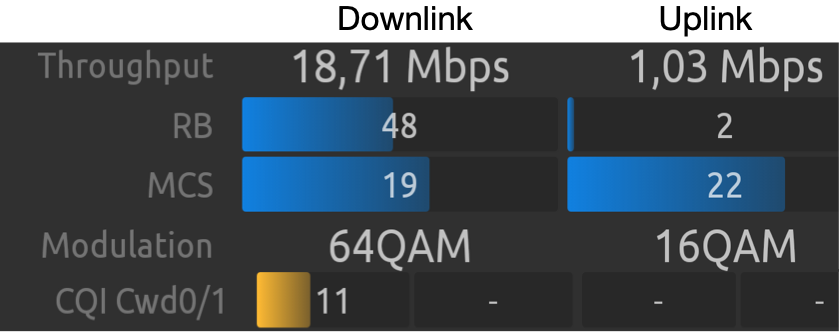}
    \caption{LTE throughput and other network parameter observed on a mobile device using Network Signaling Guru \cite{netsiguru}.}
    \label{fig:lte_resource}
\end{figure}

Figure~\ref{fig:lte_resource} shows the usage of the Modulation Scheme and the number of resource blocks for a large file download on Nexus 6 with CQI$_{11}$. LTE  supports QPSK, 16QAM, and 64QAM, i.e., each RE can carry a maximum of 2, 4, and 6 bits accordingly. Let us consider the duration of 1 RB is 1 ms ($t_s$), and there are 168 REs.  Nevertheless,   mostly 120 REs ($n$) are available for carrying data traffic. For CQI$_{11}$, the modulation scheme is 64QAM and the effective code rate $C_{CQI}=N_m/1024=0.55$. Therefore, an RE can hold only, $RE_{bits}=C_{CQI}\times M_{bits}=0.55\times6$, 3.32 bits and an RB can hold $n\times RE_{bits}=398$ bits.

The number of RBs required for a packet in a downlink can be computed using equation \ref{eq:two} by considering the additional bits for RLC and MAC headers. However, the network may not allocate the RBs according to the CQI. It may have other complex resource scheduling algorithms, as it has to deal with various types of traffic and users. The number of uplink RBs also may vary.




\section{Application}
\label{sec:codelisting}
\begin{lstlisting}[language=c, caption = {TCP sending code with/without Nagel's algorithm.}, label = {list:tcpsend}]
int val = 1;
// Disabling Nagel's Algorithm
setsockopt(sockfd,SOL_TCP,TCP_NODELAY,&one,sizeof(one));
if (connect(sockfd, &servaddr, sizeof(servaddr)) < 0)
    LOGE("[***Server Connect Error***");
for (int i = 0; i < 5000; i++) {
    usleep(20000);
    char *daat = rand_string(1300);
    gettimeofday(&tv, NULL);
    times[i] = (tv.tv_sec*1000000LL+tv.tv_usec)/1000;
    n  = write(sockfd,daat, 1300);
    if (n < 0){
        LOGE("Error sendto %s", strerror(errno));
        break;
    }
}
\end{lstlisting}

\begin{lstlisting}[language=c, caption = {TCP receiving code.}, label = {list:tcprecv}]
int BUFSIZE = 4096;
if (connect(sockfd, &servaddr, sizeof(servaddr)) < 0)
    LOGE("[***Server Connect Error***");
while (true) {
    bzero(buf, BUFSIZE);
    n = read(sockfd, buf, BUFSIZE);
    if (n > 0) {
        gettimeofday(&tvo, NULL);
        times[i]=(tvo.tv_sec*1000000LL+tvo.tv_usec)/1000;
        i = i+1;}
    else
        break;
    if (i==5000)
        break;
}
\end{lstlisting}